\documentclass{aastex}

\newcommand\gta{\lower 0.5ex\hbox{$\buildrel > \over \sim\ $}} %greater than about
\newcommand\lta{\lower 0.5ex\hbox{$\buildrel < \over \sim\ $}} %less than about
\newcommand{\kms}{\mbox{km s$^{-1}$}}

\newcommand{\Msun}{\mbox{$M_{\sun}$}}

\shortauthors{Ruiz et al.}
\shorttitle{CE-315 : Degenerate Binary}
\begin{document}

\title{CE-315 : A New Interacting Double-degenerate Binary Star.\altaffilmark{1}}

\author{Maria Teresa Ruiz\altaffilmark{2}}
\affil{Departamento de Astronom\'\i a, Universidad de Chile, Casilla 36-D, Santiago, Chile} 
\email{mtruiz@das.uchile.cl}
\author{Patricio M. Rojo}
\affil{Departamento de Astronom\'\i a, Universidad de Chile, Casilla 36-D, Santiago, Chile}
\email{projo@das.uchile.cl}
\author{Guido Garay}                         
\affil{Departamento de Astronom\'\i a, Universidad de Chile, Casilla 36-D, Santiago, Chile}
\email{ggaray@das.uchile.cl}
\and
\author{Jose Maza\altaffilmark{2}}
\affil{Departamento de Astronom\'\i a, Universidad de Chile, Casilla 36-D, Santiago, Chile}
\email{jmaza@das.uchile.cl}
\altaffiltext {1} {Based on observations obtained 
at La Silla (ESO), Chile. Project 64.H-0318.}  

\altaffiltext {2} {Visiting Astronomer, Cerro Tololo Inter American 
Observatory, NOAO, which is operated by AURA Inc., under cooperative
agreement with the National Science Foundation.}

\begin{abstract}

We present spectroscopic observations of object CE-315 revealing
a blue continuum with strong 
emission lines. Most of the detected lines are identified  with He I
or He II in emission with a handful of faint lines of nitrogen. 
Notable is the complete absence of hydrogen lines.  
The He lines exhibit triple-peaked profiles with 
remarkably broad widths, of $\sim$2000 \kms\ (FWZP).

 The observations show that 
CE-315 is an interacting binary system with an orbital period
of $65.1\pm0.7$ minutes, and a mass ratio of 0.022.
 We conclude that the most likely scenario for this
object is that of an accreting $\sim$0.77\Msun\ white dwarf with a 
$\sim$0.017 \Msun\ helium white dwarf as mass donor. 

\end{abstract}

\keywords{Stars: binaries --- stars: white dwarfs --- stars: cataclysmic
 --- stars: individual (CE-315)}  

\section{Introduction}

During the spectroscopic follow-up of proper motion stars in the Cal\'an-ESO 
Catalog (\citeauthor{ruiz01} 2001), the star CE-315 ($\mu = ~0.34'' yr^{-1}, 
~\theta = 264^{\circ}$) was found to have a spectrum consisting of a blue continuum 
with strong emission lines of He I and He II (weaker).
 The line profiles were clearly 
variable in time scales shorter than 5 minutes, with multiple components evident 
even in a low resolution discovery spectrum. 
 The spectrum of CE-315 is strikingly similar to that of the unique object 
GP Com (= G61-29), discovered by \citet{burb71}. 

\citet{nath81} found that GP Com has an orbital period of 46.5 minutes, 
and suggested that it is a double degenerate system
 in which the lower mass star 
transfers He through an accretion disk onto the higher mass component.
Evidence for CNO processed material in the accretion disk of GP Com
was found by \citet{marsh91}, who detected emission in lines of N I, 
N II, O I, Ne I and possibly Mg II. The relative strength of these lines 
indicate that nitrogen is highly over-abundant with respect to carbon and 
oxygen compared to solar values, thus providing evidence for the action of 
CNO cycle in GP Com.
The enhanced abundance of C, N, O, Mg, may have its origin in 
the thermal pulse stage of the primary star which later on transferred this 
processed material to the secondary during a common envelope phase.
More recently \citet{marsh99} published a detailed study of the kinematics
of GP Com based on time-resolved spectroscopy. He found a radial velocity
of the narrow component at the center of the triple peak He lines of
$10.8\pm1.6$ \kms\ in phase relative to the ``S''-wave (\citeauthor{nath81}
1981), consistent with an accreting white dwarf if the mass ratio of the system
$q (= M_2 / M_1)$ is $\sim 0.02$.

The understanding of the evolutionary path of a binary system leading to a 
double-degenerate (DD) system, such as GP Com,
 is a matter of considerable theoretical 
interest (\citeauthor{iben84} 1984; \citeauthor{moch89} 1989; 
\citeauthor{bran95} 1995). However, in spite of extensive
 search efforts (Robinson \& Shafter 1987; 
Bragaglia et al. 1990), there are only a handful of interacting 
binary white dwarfs known so far, and thus little is known observationally about 
this class of objects. Here we report the discovery of an accreting 
DD system with a period of 65.1 minutes, which makes it the longest period  
member of the AM CVn group.  

\section{Observations}

CE-315 was discovered during the spectroscopic follow-up of high proper motion 
stars from the Cal\'an-ESO Catalog (CE). The discovery spectrum was
obtained March 22, 1998, using the ESO 3.6m telescope at La Silla 
equipped with EFOSC2 and the B300b grism. This combination produces spectra 
with a $\sim 16$ \AA\ resolution, covering the spectral range from 3860 \AA\ to 
8070 \AA. A first inspection of the spectrum, taken with 1800s integration time,
showed no sign of any other 
lines, except for the HeI and HeII (weak) emission lines.

Shortly after the discovery, during April 3 and 4 of 1998, a series of spectra 
were obtained with the CTIO Blanco 4m telescope with the RC Spectrograph, the 
Loral 3K CCD and grating KPGL \#2. The nights were not photometric.  
However, what became clear from this series of
observations was the variable nature of the spectrum of CE-315, with
line profile changing in shape and strength in time scales of less than 300 s. 

During the nights of March 19 and 20, 1999, spectra of CE-315 were 
obtained, under photometric conditions, with the Blanco 4m telescope at CTIO. 
In this opportunity we used the RC Spectrograph with the Loral 3K
CCD detector and grating KPGL \#3, covering the spectral range between 
3750 \AA\ and 7000 \AA\ with a resolution of $\sim 4$ \AA. The 
slit was 1.5\arcsec\ wide oriented according to the parallactic angle. The 
seeing during the first night was $\sim$ 1.3\arcsec\, while during the second 
night was 1.5\arcsec. During the first night (3/19/99), 25 consecutive spectra 
of 360 s integration time each were obtained.
Taking into account the reading time, 
the time difference between each spectrum is about 510 s.
Similarly, during the second night
(3/20/99), 30 consecutive spectra of 360 s integration time were obtained. 

During March 23, 1999, using the ESO 3.6m telescope at La 
Silla equipped with EFOSC2 and grism \#16,
we obtained a sequence of 27 spectra (360 s 
integration each) with a 60 s R image taken between each spectrum. The spectral 
coverage was from 4700 \AA\ to 6755 \AA\ at a resolution of 13 \AA. The 
slit was 1\arcsec\ wide oriented according to the parallactic angle.

In all spectroscopic runs described above, we observed spectrophotometric
standards and obtained He-Ar frames in order to flux and wavelength
calibrate the data using IRAF packages.

Photometry of CE-315 was obtained in April 12, 1999, at the CTIO 0.9m
telescope using a CCD (Tek 2k \#3), under photometric conditions. The following
magnitudes were obtained :~ B=17.23, V=17.67, R=17.47, and I=17.27 
(Kron-Coussins). The series of R images taken with the 3.6m telescope,
described above, clearly showed that the object is photometrically variable
in time scales of minutes, therefore, given that the
integration times used for obtaining
the photometry with the 0.9m were 1200 s in B, 900 s in V, and 660 s in R and I,
the above magnitudes are only an average. The observed variations in R
occurred in time scales of two to three minutes
and with amplitudes up to 0.2 mag.

In March 21, 1999 using ESO 3.6m telescope and EFOSC2 a direct He I 
(6678~ \AA) image of CE-315 was obtained, with an integration time of 1200 s. 
The PSF shape and size of CE-315 was similar to the rest of the stars in
the frame indicating that, as in GP Com (Stover, 1983),
there is no resolved He emission surrounding CE-315 that could be attributed
to mass ejections during previous evolutionary phases. 

\section{Results}

\subsection{Average Spectra}

Figure 1 presents the average of all the individual spectra taken 
in a given observing run.
The upper panels show the average spectrum
obtained from the CTIO observations made during April 4, 1998, covering the 
spectral range from 3750 \AA\ to 8830 \AA. The lower panels 
show the average spectra obtained from the observations made in March 1999    
at CTIO and ESO, respectively. Most of the detected lines 
 can be identified with either He I or He II lines 
in emission, with a handful of faint lines of N I. Notable is the 
complete absence of hydrogen emission lines.

\subsection{Line Profiles}

 He I lines from CE-315 show triple-peaked profiles, although in a few
cases the emission from the central component almost completely
dominates the spectrum. This is illustrated in Figure 2, which shows the 
profile of the $\lambda$4471.5, $\lambda$5875.6, and $\lambda$6678.1 lines 
of He I. The relative intensities of the outer
and central components differ from
line to line, implying that the line profiles are made up of emission
that originates from two independent sources. We suggest that the emission 
from the outer components of the line profiles arises from an accreting disk 
(double-peaked profiles are characteristics of disk emission), whereas the 
central component has a different origin. 

The profiles of the He I emission are remarkably broad, covering a
velocity range of about 2000 \kms\ (FWZP). FWHM line widths, determined by
simultaneously fitting 3 Gaussian components to the observed profiles, are
typically 500 \kms\ for the blue disk feature, 580 \kms\ for the red disk 
feature, and 345 \kms\ for the central feature.

The above characteristics suggest that CE-315 is a binary system with the
lower mass star transferring mass by way of an accretion disk. In what follows
we present the determination of observed parameters of this system.

\subsection{Orbital Period}

For the determination of the orbital period of the system we followed the method
devised by \citet{nath81}, in which the profile of each line is
divided in a blue wing region and a red wing region, and the difference
between the integrated intensity in the two bands is analyzed for periodicity.

Figure 3 shows the difference in red and blue integrated intensities (R-B)
versus time for three of the most intense lines within the observed spectral
range and for three different nights.
These non-uniformly spaced time series were
investigated using the periodogram analysis for unevenly spaced data devised 
by Scargle (1982), and developed in \citet{press92}. The resulting power spectrum are also shown in Figure 3.
The periods determined using the data taken on March 19, March 20, and March 24 
of 1999 are, respectively, $65.2\pm0.9$ minutes, $64.4\pm1.1$ minutes, and 
$65.7\pm1.2$ minutes. In what follows we adopt a period for the system of 
65.1 minutes.

\subsection{The ``S'' Wave}

An inspection of the line profiles shows that the dominant form of the
modulation is due to changes in intensity of the line profiles. To study
this modulation in more detail we subtracted from each individual spectrum
an average spectrum and analyzed the resulting spectra as a function of
phase. Figure 4 presents trailed spectra, using the data
obtained during March of 1999 at CTIO, showing that
most of the modulation is due to an emission feature that wanders in
velocity between the blue and red velocity of the disk component. Due to 
its appearance on single trailed spectra, this feature is usually labeled as
the ``S'' wave. The ``S'' feature is thought to mark a bright region of
enhanced emission where the gas stream from the donor hits the
accretion disk. In the reference frame rotating with the binary system
the bright spot is fixed in position, but the velocity of the gas in the spot
is similar to the velocity of the disk at that position.
The velocity variation shown in Figure 4 is well fitted by a sinusoidal function, 
with a semiamplitude of $630\pm15$ \kms. This suggests that the hot spot is 
located at a radius of about 0.7 times the radius of the outer edge of the disk 
and that it is undergoing circular orbital motions.
Similar to what is observed in GP Com, in CE-315
 the intensity of the ``S'' wave 
is noticeably weaker when moving from blueshifted to redshifted velocities, which 
is probably due to effects of the inclination of the system.

\subsection{Disk velocity and Orbital Velocity of the Primary Component}

The projected rotational disk velocity, $V_d~ sin i$, can be determined from
the half-separation of the outer peaks. From the profiles in the 
$\lambda$5875.6 and $\lambda$6678.1 lines observed with the CTIO telescope during 
March 1999, which are the lines with the largest velocity
resolution, we derive a half-separation of $510\pm10$ \kms.

Assuming a simple model of a Keplerian disk rotating around the primary
component of a binary system, the radial velocities of the red and blue
components of the disk profiles, $V_{r}$ and $V_{b}$, respectively, should
vary with phase, $\phi$, as (e.g., \citeauthor{sma76} 1976)
\begin{equation}
 V_{r,b} = \gamma \pm V_d~ sin i + K_1~ sin \phi ~~,
\end{equation}
where $\gamma$ is the systemic velocity
and $K_1$ is the amplitude of the radial velocity of the primary
component. 
For each line we computed $V_{r,b}$ from the observed profile using the
expression
\begin{equation}
 V_{r,b} = {{\int_{r_1,b_1}^{r_2,b_2} I(v) v dv}\over{\int_{r_1,b_1}^{r_2,b_2} 
I(v) dv}}~~,
\end{equation}
where $I(v)$ is the observed intensity at velocity $v$, and $r_1,b_1$ and
$r_2,b_2$ are the initial and final velocities of the range of 
red and blue disk velocities, respectively.
Figure 5 shows the dependence of the blue disk velocity with phase 
derived using the data taken during the night of March 20 of 1999. 
The blue disk velocity corresponds to an average of the individual $V_b$ 
determined for the $\lambda$5875.6 and $\lambda$6678.1 lines. 
A least squares fit to the observed trend gives
$\gamma-V_d~ sin i= -490$ \kms\ and $K_1=8\pm1$ \kms.
The first order velocity moment used by us
as representative of the left hand side of eqn.(1), should be less sensitive
than other estimates to the presence of perturbations such as the ``S'' wave.

\section{Discussion}

\subsection{The Binary System}

Given the short period of CE-315 (65.1 minutes), two candidates for the
donor star are possible (see \citeauthor{fau72} 1972) ~:~ i) a helium ZAMS of
$\sim1.7$\Msun, and ~ii) a degenerate helium low-mass white dwarf with
M$_2$ $\sim 0.017$ \Msun\ (\citeauthor{war95} 1995). The first alternative 
can be easily dismissed, since in that case the light of the ZAMS 
star would dominate the spectrum. We 
therefore conclude that the donor is a low-mass helium degenerate.
In addition, the large rotational velocities of the
accretion disk, of the order of 510 \kms, require that the accreting star
be also a collapsed star. Since CE-315 is not known to be a
strong X-ray source, the accreting star is unlikely to be a
neutron star or a black hole, and therefore we conclude it must be a white
dwarf.

From the observed projected rotational velocity of the disk and the orbital
velocity of the primary component, it is possible to compute the mass ratio of 
the binary system, $q= {{m_2}\over{m_1}}$,
using the expression (\citeauthor{sma76} 1976)
\begin{equation}
 {{q}\over{(1+q)^{1/2}}} = \left({{K_1}\over{V_d sin i}}\right)
   \left({{a}\over{r_d}}\right)^{1/2}~~,
\end{equation}
where $a$ is the separation of the system and $r_d$ the disk
radius. Using $V_d~ sin i$= 510 \kms, $K_1$=8 \kms, and assuming
${{r_d}\over{a}}= 0.5$, we obtain $q=0.022$. The last assumption is based 
on the computations of \citet{lin76}, who found that the accretion 
disk is comparable in size to the Roche lobe of the primary.
Assuming M$_2$ = 0.017 \Msun, the derived mass ratio
implies that the accretor mass is $\sim0.77$ \Msun.

\subsection{Origin of the Central Component}

The emission from the central component of CE-315
is barely resolved, exhibiting a line width of 
$\sim$350 \kms, broader than that of the central feature in GP 
Com which is $\sim$ 120 \kms\ (\citeauthor{marsh99} 1999).
 The ratio of intensities 
of the central to the disk component differ from line to line, increasing for 
higher excitation lines. In particular, in the He II 4686 line 
the emission from the central component completely dominates over that 
of the disk (see Table 1).
These results suggest that 
the emission from the central component arises from a region
with a temperature higher than that in the 
disk, high enough to ionize helium.

Our direct He I ($\lambda6678$) image of CE-315 shows that the helium emission 
is unresolved, thus it is unlikely that the emission from the central component 
arises in a nebular shell, produced by mass ejections during a previous 
evolutionary phase. Further, the 
small radial velocity variation of the central feature makes it unlikely that 
it arises in the mass donor. 
Thus, we suggest that the central emission originates near the accreting white 
dwarf. Possible sites are: i) the inner boundary layer produced by accretion 
onto the white dwarf, and ii) the region of inflow from the Alfven surface to 
the star for a magnetized white dwarf (see \citeauthor{fran92} 1992). 
In the first case the temperature of the boundary layer is about 6 times larger 
than that of the disk, and implies that CE-315 should be a soft X-ray source.
An interesting difference between CE-315 and GP Com is that in CE-315 the
radial velocity of the central component is similar to the mean radial velocity
of the double peaks, whereas in GP Com the central component
is significantly redshifted with respect to the later.

\subsection{Evolutionary Stage of CE-315} 

Following the same line of argument that lead 
\citet{nath81} to conclude that GP Com is an 
evolutionary descendant of a system like AM CVn, we propose that CE-315,
having an almost pure He emission line spectrum and a longer period,
is also a descendant of an AM CVn system but at a slightly more evolved 
stage than GP Com.

The number of double degenerate (DD) systems detected in different systematic
searches (see \citeauthor{robi87} 1987, \citeauthor{foss91} 1991,
\citeauthor{brag90} 1990)
have turned out to be very small. In an effort to explain this observational
fact, \citet{hern97} estimated the expected population of DD systems in
the solar neighborhood taking into account the scale height of these objects
and the white dwarf's cooling time. As a result they find that the predictions
from the models are in agreement with the observed scarcity of DD systems. 
In particular, they predict a number density of DD systems with a period of
65 minutes (like CE-315) between $10^{-5}$ and $1.4\times 10^{-5}$ pc$^{-3}$
depending on whether the stellar formation rate is compatible with the models 
of chemical evolution of the Galaxy or with
the white dwarf's luminosity function.
Unfortunately the distance to CE-315 is not known, which prevents us
from deriving a density to compare with model predictions. 
An accurate determination of the distance to CE-315 (and GP Com)
through trigonometric parallaxes, would allow us to calculate the number
density of DD systems in the CE survey, with well defined limits
in $m_R  \sim 19.5$ and $\mu = 0.2\arcsec$ yr$^{-1}$, which would be crucial to 
test model predictions, thus making an important contribution to 
a better understanding of the evolutionary phase of this class of objects.

\acknowledgements

M.T.R. and P.R. received partial support from FONDECYT grant 1980659 and 
Catedra Presidencial en Ciencias. G.G. acknowledges support from a 
Chilean Presidential Science Fellowship and FONDECYT project 1980660.
%

%TABLE1
 \clearpage

\begin{deluxetable}{p{1.0cm}ccc}
\tablecolumns{4}
\tablewidth{0pt}
\tablecaption{Emission Line Intensities}
\tablehead{
\colhead{Element}  &
\colhead{$\lambda$} &
\multicolumn{2}{c}{$F_\lambda$($ergs~ cm^{-2} ~s^{-1}$)} \\
\colhead{} &
\colhead{($\AA$)} &
\colhead{Disk}  &
\colhead{Central}}

\startdata
HeI & 3819.6 & 2.5229e-15 & 2.770E-15 \\ 
HeI & 3888.6 & 9.409e-15 & 6.218E-15 \\ 
HeI & 3964.7 & 1.0902e-15 & 1.465E-15 \\ 
HeI & 4026.0 & 5.5893e-15 & 3.685E-15 \\ 
HeI & 4388.0 & 1.2477e-15 & 1.895E-15 \\ 
HeI & 4471.5 & 5.472e-15 & 7.201E-15 \\ 
HeI & 4713.1 & 2.328e-15 & 1.175E-15 \\ 
HeI & 4921.9 & 3.44e-15 & 2.687E-15 \\ 
HeI & 5015.7 & 6.653e-15 & 2.240E-15 \\ 
HeI & 5875.6 & 1.5886e-14 & 4.929E-15 \\ 
HeI & 6678.1 & 8.434e-15 & 2.617E-15 \\
HeII & 4685.7 & 9.529e-16 & 1.015E-15 \\ 
\enddata

\end{deluxetable}

\clearpage
\figcaption[f1.eps] {Average of individual spectra taken during each
observing run. The upper panels show the spectrum obtained with the
4m telescope at CTIO in April 1998, covering the spectral range
from 3750 \AA\ to 8830 \AA (the spike at $\lambda 4388~ \AA$ is a
defect in the CCD).
The lower panels show the average spectra
obtained in March 1999 with the CTIO 4m and ESO 3.6m telescopes, 
respectively.\label{fg:figure1}}
 
\figcaption[f2.eps] {
Profiles of the $\lambda$4471.5, $\lambda$5875.6 and $\lambda$6678.1
lines. The difference in the relative intensities between the
central component and the disk component between the $\lambda$4471.5 line and 
the other lines is evident. 
\label{fg:figure2}}

\figcaption[f3.eps] {Left panels: Red minus Blue (R-B) integrated 
 line intensities  versus time for three of the most intense lines 
 (dotted: $\lambda$6678.1, dashed: $\lambda$5875.6,  
 dot-dashed: $\lambda$5015.7) and for three different nights.
Right panels: Power spectrum obtained using a periodogram analysis
of the data in the left panels.
The derived periods are 65.2 $\pm$ 0.9 minutes (3/19/99),
64.4 $\pm$ 1.1 minutes (3/20/99) and 65.7 $\pm$ 1.2 minutes (3/24/99).     
\label{fg:figure3}}

\figcaption[f4.eps] {Trailed spectra of the CTIO 1999 observations   
phase-folded and with the same cycle repeated three times.  A lowest-flux
spectrum has been subtracted. The intensity is displayed in
a logarithmic scale. This figure shows that most of the modulation observed
in figure 3 is due to a feature that wanders in velocity between the
blue and red velocity of the disk component, called the ``S'' wave.
\label{fg:figure4}}

\figcaption[f5.eps] {
Radial velocity of the blue disk component, $V_b$, (average of the 
$\lambda$5875.6 and $\lambda$6678.1 lines) versus phase. The fit indicates 
a sinusoidal with an amplitude of 8 \kms. 
\label{fg:figure5}}

\clearpage

\begin{figure}[p]
\plotone{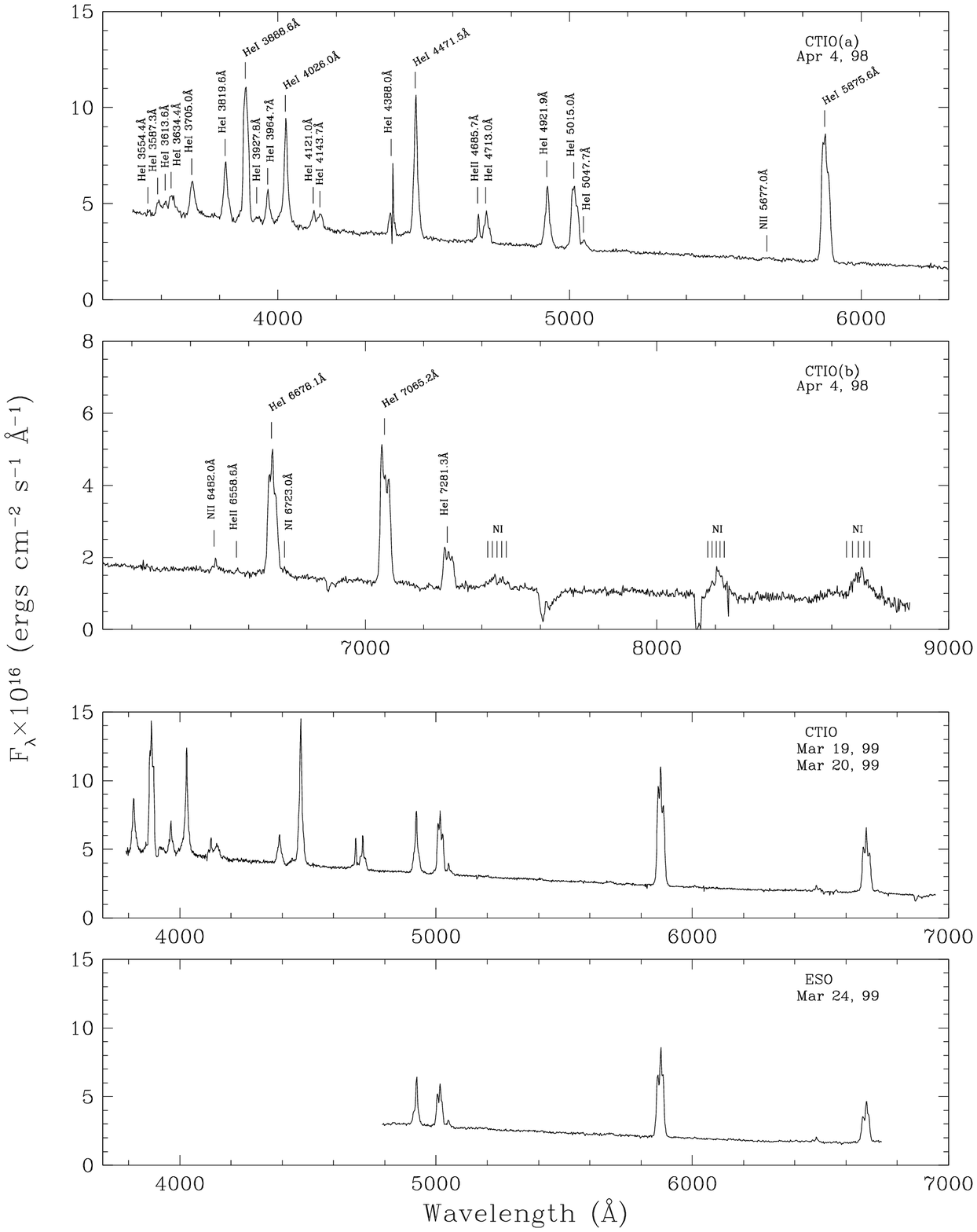}
\begin{flushright}
Figure \ref{fg:figure1}
\end{flushright}
\end{figure}

\begin{figure}[p]
\plotone{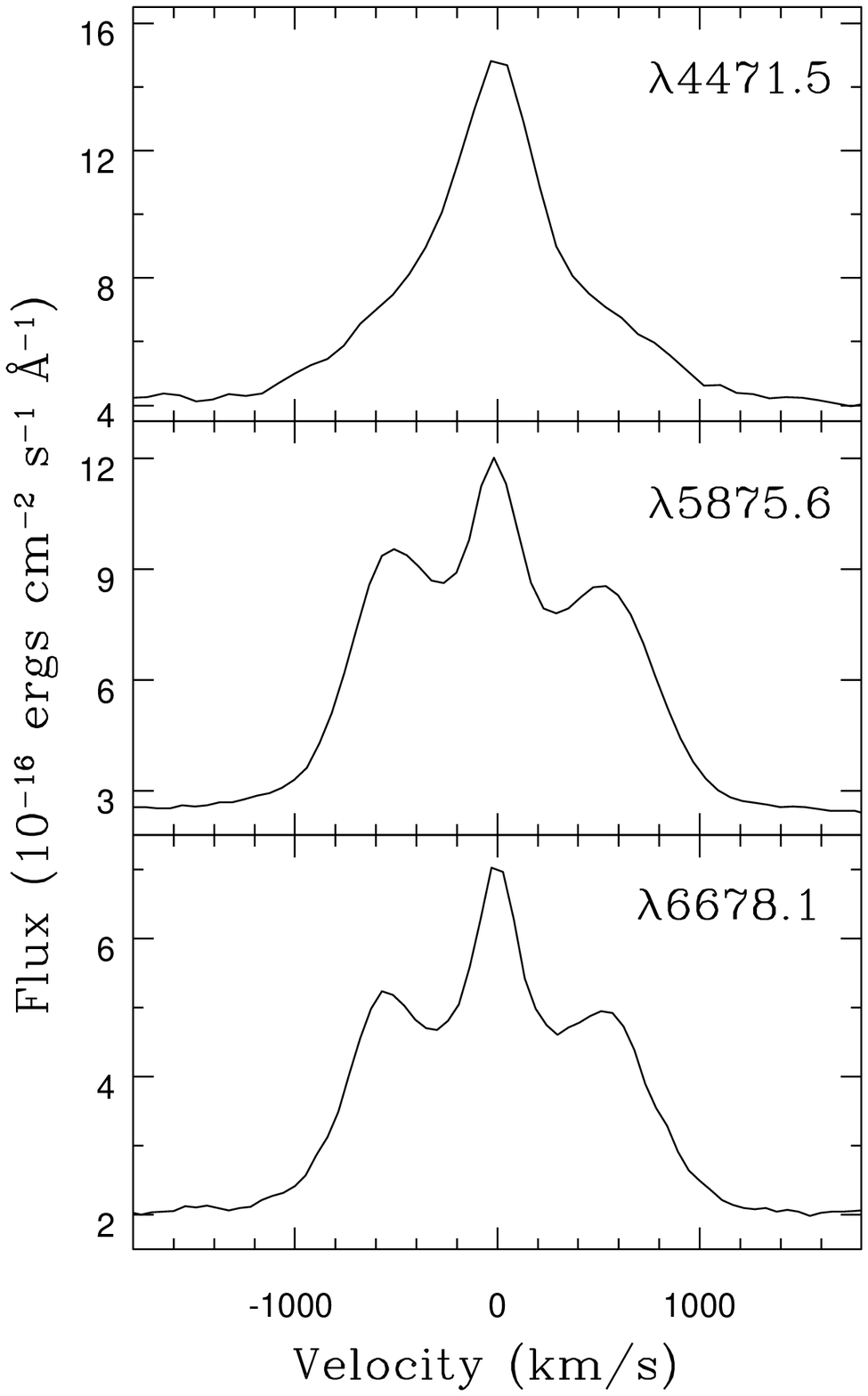}
\begin{flushright}
Figure \ref{fg:figure2}
\end{flushright}
\end{figure}

\begin{figure}[p]
\plotone{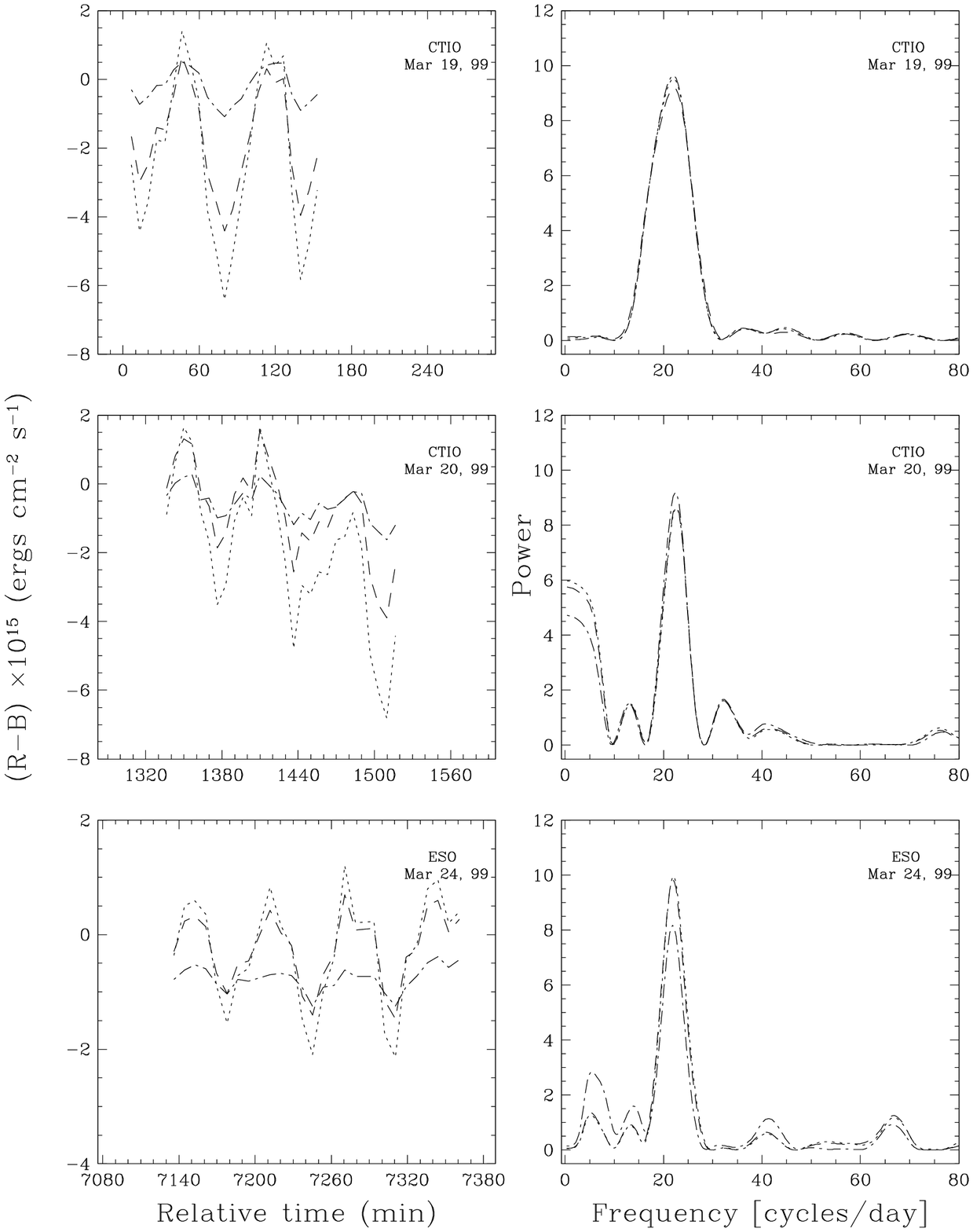}
\begin{flushright}
Figure \ref{fg:figure3}
\end{flushright}
\end{figure}

\begin{figure}[p]
\plotone{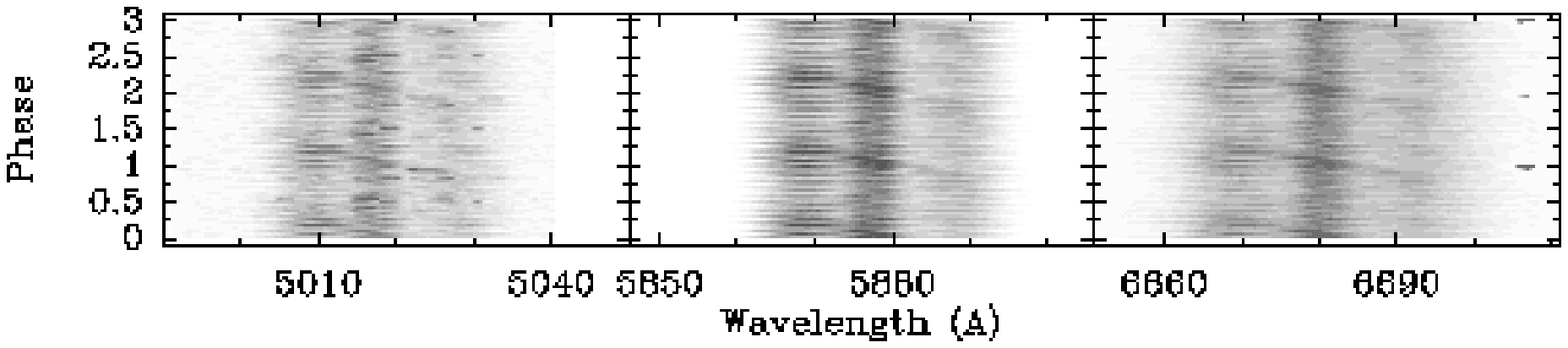}
\begin{flushright}
Figure \ref{fg:figure4}
\end{flushright}
\end{figure}

\begin{figure}[p]
\plotone{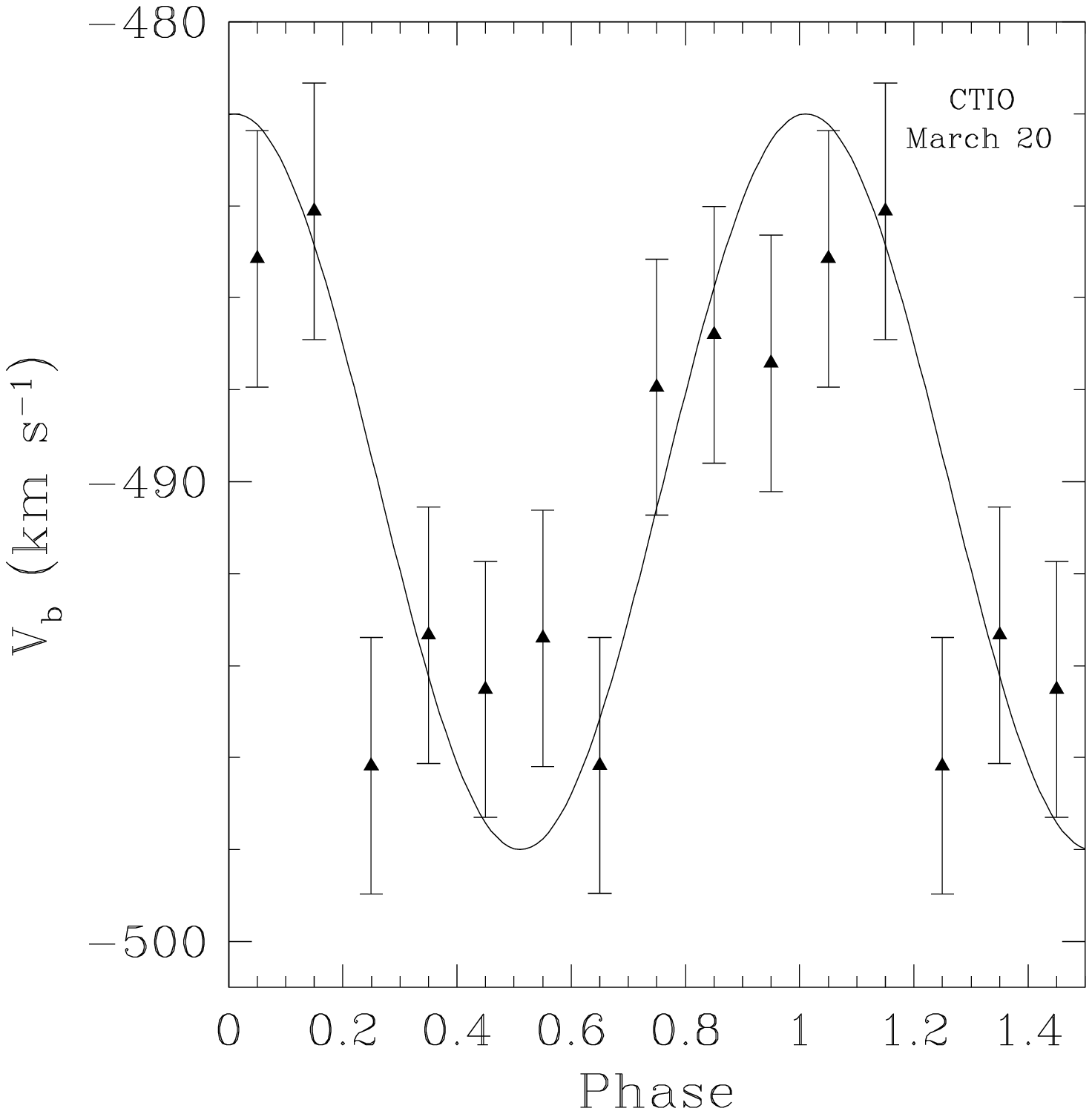}
\begin{flushright}
Figure \ref{fg:figure5}
\end{flushright}
\end{figure}

\end{document}